\documentclass[twocolumn,showpacs,preprintnumbers,amsmath,amssymb,prb,superscriptaddress]{revtex4}

\usepackage{epsfig}
\usepackage{graphicx}
\usepackage{dcolumn}
\usepackage{bm}
\usepackage{color} 

\begin{document}

\title{
Magnetic Field-Induced Superconductor-Insulator-Metal Transition in an Organic Conductor: An Infrared Magneto-Optical Imaging Spectroscopy
}

\author{Tatsuhiko Nishi}
\altaffiliation{Present address: Department of Chemistry, Nagoya University, Furo-cho, Chikusa-ku, Nagoya 464-8602, Japan}
\affiliation{School of Physical Sciences, The Graduate University for Advanced Studies (SOKENDAI), Okazaki 444-8585, Japan}
\author{Shin-ichi Kimura}
 \altaffiliation[Electronic address: ]{kimura@ims.ac.jp}
\affiliation{UVSOR Facility, Institute for Molecular Science, Okazaki 444-8585, Japan}
\affiliation{School of Physical Sciences, The Graduate University for Advanced Studies (SOKENDAI), Okazaki 444-8585, Japan}
\author{Toshiharu Takahashi}
\affiliation{Research Reactor Institute, Kyoto University, Osaka 590-049, Japan}
\author{Hojun Im}
\affiliation{School of Physical Sciences, The Graduate University for Advanced Studies (SOKENDAI), Okazaki 444-8585, Japan}
\author{Yong-seung Kwon}
\affiliation{Department of Physics and Institute of Basic Science, Sungkyunkwan University, Suwon 440-746, South Korea}
\author{Takahiro Ito}
\affiliation{UVSOR Facility, Institute for Molecular Science, Okazaki 444-8585, Japan}
\affiliation{School of Physical Sciences, The Graduate University for Advanced Studies (SOKENDAI), Okazaki 444-8585, Japan}
\author{Kazuya Miyagawa}
\affiliation{Department of Applied Physics, The University of Tokyo, Hongo, Bunkyo-ku, Tokyo 113-8656, and CREST, JST, Kawaguchi Saitama 332-0012, Japan}
\author{Hiromi Taniguchi}
\affiliation{Department of Physics, Saitama University, Saitama 338-857, Japan}
\author{Atsushi Kawamoto}
\affiliation{Department of Physics, Hokkaido University, Sapporo 060-0810, Japan}
\author{Kazushi Kanoda}
\affiliation{Department of Applied Physics, The University of Tokyo, Hongo, Bunkyo-ku, Tokyo 113-8656, and CREST, JST, Kawaguchi Saitama 332-0012, Japan}

\date{\today}

\begin{abstract} 
The magnetic field-induced superconductor--insulator--metal transition (SIMT) in partially deuterated $\kappa$-(BEDT-TTF)$_{2}$Cu[N(CN)$_{2}$]Br, which is just on the Mott boundary, has been observed using the infrared magneto-optical imaging spectroscopy.
The infrared reflectivity image on the sample surface revealed that the metallic (or superconducting) and insulating phases coexist and they have different magnetic field dependences.
One of the magnetic field dependence is SIMT that appeared on part of the sample surface.
The SIMT was concluded to originate from the balance of the inhomogenity in the sample itself and the disorder of the ethylene end groups resulting from fast cooling.
\end{abstract}


\maketitle

\section{Introduction}
A family of $\kappa$-(BEDT-TTF)$_{2}X$ (BEDT-TTF = bis(ethylenedithio)-tetrathiafulvalene, $X$~=~Cu[N(CN)$_{2}$]Br, Cu[N(CN)$_{2}$]Cl, Cu(NCS)$_{2}$, etc.) salts is a charge-transfer group with quasi-two-dimensional electronic properties.~\cite{Ishiguro1998}
These salts have attractive physical properties, including a range of antiferromagnetic insulating (AFI) to superconducting (SC) ground states depending on the pressure,~\cite{Kagawa} deuteration,~\cite{Kawamoto1997} cooling rate,~\cite{Taniguchi1999} and magnetic field strength.~\cite{Taniguchi2003}
In these salts, the $\kappa$-(BEDT-TTF)$_{2}$Cu[N(CN)$_{2}$]Br (abbreviated $d[0,0]$ hereafter) is in the SC state in the vicinity of a SC--AFI boundary, named the Mott boundary.~\cite{Kanoda1997,Kawamoto1998}
Additionally, a perfectly deuterated salt ($d[4,4]$) formed under slow cooling conditions is believed to exist on the Mott boundary.
In the case of the fast cooling conditions, the ground state changes to be in the AFI phase because the disorder originating from the mixing of the eclipsed and staggered conformations of the ethylene end groups in the BEDT-TTF molecules of the sample affects the ground state characteristics.~\cite{Geiser1991}
In intermediate materials in which the deuterium is randomly distributed in the ethylene groups, the ground state is on the boundary between the SC and AFI phases.
In these cases, a partial SC state in the material is observed.~\cite{Taniguchi1999}
For instance, a 75~\%-deuterated material ($d[3,3]$) that exhibits a uniform metallic (or SC) state under slow cooling conditions changes to an insulating state with fast cooling conditions.~\cite{Taniguchi1999}
The spatial distribution of the center of the spectral weight ($\langle\omega\rangle$) of the reflectivity spectra [$R(\omega)$] in the 10~$\mu$m-scale region on the sample surface indicates that the inhomogenity contained in the $d[3,3]$ sample itself is the origin of the metal--insulator phase separation.~\cite{Nishi2005}
In the case of the presence of magnetic fields, the 50~\%-deuterated $\kappa$-(BEDT-TTF)$_{2}$Cu[N(CN)$_{2}$]Br ($d[2,2]$) under fast cooling conditions directly changes from the SC to AFI states in spite of the fact that the SC state of the material formed with slow cooling conditions normally changes to a paramagnetic metallic (PM) state with increasing magnetic field strength.~\cite{Taniguchi2003}

In this paper, we report the magnetic field dependence of the spatial image of the SC--AFI transition in $d[2,2]$ using the infrared magneto-optical imaging spectroscopy.
In addition, we found a magnetic field-induced SC--AFI--metal transition (SIMT) in a tiny region on the sample surface that appeared under the appropriate conditions of both the inhomogenity and disorder of the ethylene groups caused by fast cooling conditions.
The origin of the SIMT and the conceptual phase diagram of the magnetic field ($B$) to $U/W$ in which $U$ and $W$ are the intradimer Coulomb repulsion energy and the band width, respectively, of $d[2,2]$ at 5~K are also discussed.

\section{Experimental}
Single crystals of $d[2,2]$ were grown using a conventional electrochemical oxidation method.
The unpolarized infrared reflectivity spectra and the spatial image were measured at the infrared magneto-optics station of the beam line 43IR of a synchrotron radiation facility, SPring-8.~\cite{Kimura2001,Kimura2003}
To observe the cooling rate dependence of the samples, two cooling rates, -17~K/min for fast cooling, and -0.05~K/min for slow cooling, were used in the temperature range from 90 to 70~K.
To check the reproducibility of the spectra and the spatial images, the same fast cooling experiment was repeated, and also another $d[2,2]$ sample was tested.
A strong reproducibility was obtained, indicating that the $R(\omega)$ and the spatial image mainly originate from the inhomogenity of the sample itself, which has been demonstrated in previous studies.~\cite{Nishi2005}
The infrared $R(\omega)$ imaging was measured in the wave number range from 870 to 8,000~cm$^{-1}$ at a temperature of 5~K and magnetic fields of 0, 5, and 10~T.
The measurements were done only with increasing magnetic field strength only.

\begin{figure}[t]
\begin{center}
\includegraphics[width=6cm]{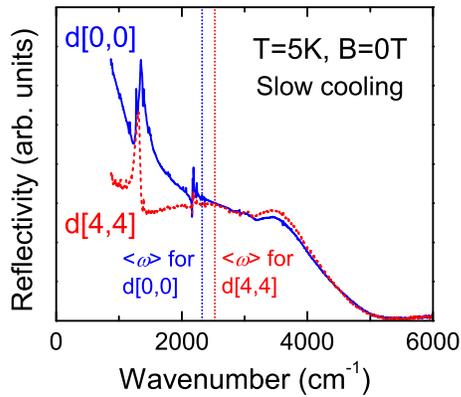}
\end{center}
\caption{
(Color online)
Typical reflectivity spectra [$R(\omega)$] in the superconducting ($d[0,0]$, $T$~=~5~K, $B$~=~0~T) and antiferromagnetic insulating ($d[4,4]$, $T$~=~5~K, $B$~=~0~T) phases and their centers of spectral weight ($\langle\omega\rangle$).
}
\label{fig1}
\end{figure}
To acquire the spatial imaging data, a total of 931 spectra were acquired in a 360$\times$360~$\mu$m$^{2}$ region at steps of 12~$\mu$m with a spatial resolution of less than 20~$\mu$m.
The spatial image of the reflectivity spectra were plotted using $\langle\omega\rangle$ obtained from the following function:
\[
\langle\omega\rangle = \int_{\omega_1}^{\omega_2} \omega R(\omega) d\omega/ \int_{\omega_1}^{\omega_2} R(\omega) d\omega
\]
for each $R(\omega)$.~\cite{Nishi2005}
Here, $\omega_1$ is the lowest accessible wavenumber, 870~cm$^{-1}$, and $\omega_{2}$ is set to 5,000~cm$^{-1}$, above which there is no difference between the metallic and non-metallic spectra.~\cite{Gri1999}
The reason for this convergence in the spectra is that the change in the electronic structure due to the Mott transition~\cite{DMFT1993} results in a shift of the optical spectral weight, because the optical spectra indicate the relative energy difference between the occupied and unoccupied states.
For instance, typical $R(\omega)$ spectra in the SC ($d[0,0]$, $T$~=~5~K, $B$~=~0~T) and AFI ($d[4,4]$, $T$~=~5~K, $B$~=~0~T) phases and their respective $\langle\omega\rangle$ values are shown in Fig.~\ref{fig1}.
Although the origin of the spectral shape between 2,000 and 4,000~cm$^{-1}$ has not been concluded yet, the difference in the electronic structure is directly reflected in $R(\omega)$ as well as $\langle\omega\rangle$.
The metal--insulator character in $R(\omega)$ mainly appears below 1000~cm$^{-1}$ as shown in Fig.~\ref{fig1}.
However the spatial resolution becomes poorer than the present method because of the diffraction effect of light.
Therefore, the $\langle\omega\rangle$ was employed as being the representative of the shape of $R(\omega)$ as well as the electronic structure.

To check the wavenumber distribution of $\langle\omega\rangle$, the pixel number of $\langle\omega\rangle$ as a function of $\langle\omega\rangle$ was also derived.
From our previous work, the rough boundary ($\omega_{MI}$) between metallic and insulating characteristics in the same analysis method was evaluated to be around 2,350~cm$^{-1}$.~\cite{Nishi2005}
The other analysis methods, the reflectivity ratio between at the lowest accessible wavenumber of 870~cm$^{-1}$ and at the shoulder structure of 3,500~cm$^{-1}$ and the wavenumber shift of the $\nu_{3} (a_{g})$ molecular vibration mode,~\cite{Sasaki2004} gave the similar result.
Note that this method cannot separate an SC state from a normal metallic state because SC behavior does not effect $R(\omega)$ down in the accessible lowest wavenumber of 870~cm$^{-1}$.
Then the obtained metallic $R(\omega)$ below the upper critical field ($H_{c2}\sim$4~T) and below the critical temperature ($T_{c}\sim$11~K) is regarded to be in an SC state.

\section{Results and Analysis}
\begin{figure}[t]
\begin{center}
\includegraphics[width=7.5cm]{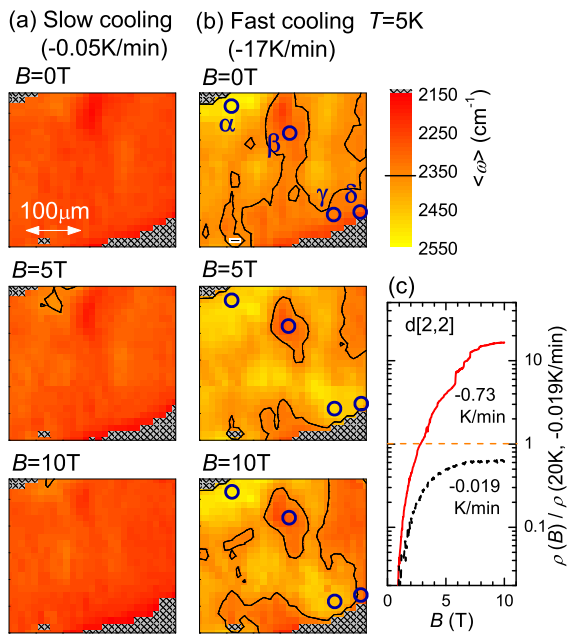}
\end{center}
\caption{
(Color)
Magnetic field and cooling rate (slow cooling in (a) and fast cooling in (b)) dependencies of the spatial image of the center of the spectral weight ($\langle\omega\rangle$) of 50 \%-deuterated $\kappa$-(BEDT-TTF)$_{2}$Cu[N(CN)$_{2}$]Br ($d[2,2]$) at $T$~=~5~K.
The blue circles in (b) indicate the points of the different magnetic field dependencies on the sample surface are shown in Fig.~\ref{fig4}.
The black lines indicate the rough M--I boundary ($\omega_{MI}$) of 2,350~cm$^{-1}$ and the lower and higher wavenumbers indicate the insulating and metallic (superconducting) reflectivity spectra, respectively.
The hatched area is the outside of the sample.
(c) Magnetic field and cooling rate dependences of the normalized resistivity of $d[2,2]$ at 5.5~K for the reference.
Though the cooling rate of -0.73~K/min is different from that of (b), the physical character is same as shown in Ref. 5.
}
\label{fig2}
\end{figure}
\begin{figure}[t]
\begin{center}
\includegraphics[width=7.5cm]{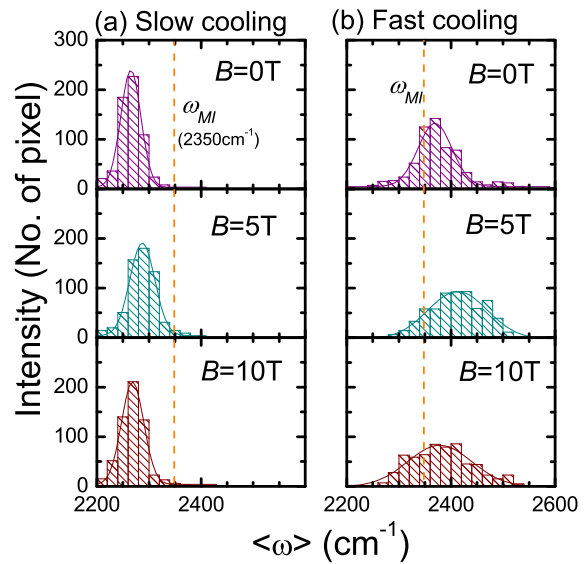}
\end{center}
\caption{
(Color online)
Statistical distributions of $\langle\omega\rangle$ for $d[2,2]$ derived from Fig.~\ref{fig2}.
The Gaussian fittings to the histograms are also plotted using lines.
The fitting results are discussed in the text.
}
\label{fig3}
\end{figure}
The spatial images of $\langle\omega\rangle$ obtained under the slow cooling (a) and fast cooling (b) conditions are shown in Fig.~\ref{fig2}.
In the case of slow cooling at 0~T, the spatial image is nearly monochromatic, and then the $\langle\omega\rangle$ value over the whole sample surface is lower than $\omega_{MI}$.
This indicates the whole sample surface is in the metallic state.
The metallic state does not change with increasing magnetic field strength up to 10~T.
These results are consistent with the electrical resistivity data under magnetic fields as shown in Fig.~\ref{fig2}c.~\cite{Taniguchi1999,Taniguchi2003}

In the case of the fast cooling condition at 0~T, $\langle\omega\rangle$ shifts to the higher wavenumber side.
This indicates that the insulating region expands compared with the results obtained under the slow cooling condition.
The distribution of $\langle\omega\rangle$ values shown in Fig.~\ref{fig3}b also crosses $\omega_{MI}$, \textit{i.e.}, the insulating state mixes with the metallic (or SC) state at the sample surface.
With increasing magnetic field strength up to 5~T, $\langle\omega\rangle$ shifts to the higher wavenumber side and the insulating region expands in contrast to the constant $\langle\omega\rangle$ distribution observed in the slow cooling experiment.
This result is consistent with the electrical resistivity data if the following explanation is accurate.
At 0~T, the SC domains connect to one another resulting in the electrical resistivity dropping to zero even if the AFI domains exist.
Actually, the percolation of the SC state appears in the top figure of Fig.~\ref{fig2}b.
The coexistence of SC and AFI states is also consistent with the ac-susceptibility data that does not show perfect diamagnetism.~\cite{Taniguchi1999}
When magnetic fields are applied, the AFI domains expand and the SC (or metallic) domains disconnect as shown in the middle figure of Fig.~\ref{fig2}b.
As a result, the electrical resistivity drastically changes to reflect an insulating state.
In fact, the re-entrant SC phase appears at the boundary of the AFI--SC transition in $\kappa$-(BEDT-TTF)$_{2}$Cu[N(CN)$_{2}$]Cl (denoted $\kappa$-Cl hereafter)~\cite{Ito1996} and $d[4,4]$~\cite{Ito2000} with increasing pressure.
The origin of this observation has been revealed to be a mixture of the SC and AFI domains resulting from the pressure dependent electrical resistivity of $\kappa$-Cl~\cite{Kagawa} and from the pressure dependent infrared $R(\omega)$ of $d[4,4]$.~\cite{Kimura2007}
Therefore, the S--AFI transition observed in the electrical resistivity measurement originates from the increase in the size of the AFI region in the sample.
The coexistence of SC and AFI domains has been studied using the half-filled Hubbard model in the limit of infinite dimensions.~\cite{Laloux1994}
This paper pointed out that the magnetic field-induced metal--insulator transition occurs in the vicinity of $U/W\sim1$.
Therefore, the origin of the SC--AFI transition in $d[2,2]$ might be the predicted metal--insulator transition in which the critical magnetic field of the Mott transition is below $H_{c2}$.

\begin{figure}[t]
\begin{center}
\includegraphics[width=7cm]{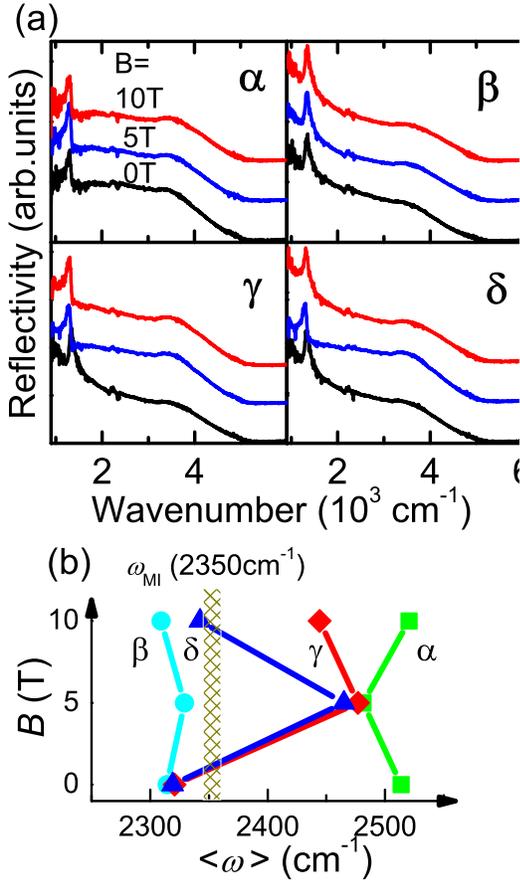}
\end{center}
\caption{
(Color online)
The magnetic field dependencies of the $R(\omega)$ spectra (a) and of $\langle\omega\rangle$ (b) at points $\alpha, \beta, \gamma$, and $\delta$ in Fig.~\ref{fig2}.
The $R(\omega)$ spectra at points $\alpha$ and $\beta$ exhibit the magnetic field-independent insulating (I) and metallic (M) characteristics, respectively.
On the other hand, the spectra at points $\gamma$ and $\delta$ change from M (or SC)$\rightarrow$I and M (or SC)$\rightarrow$I$\rightarrow$M, respectively, with increasing magnetic field.
}
\label{fig4}
\end{figure}
With increasing magnetic field strength up to 10~T, part of the sample was observed to change from an insulating to metallic state as shown in Fig~\ref{fig2}b.
However, the metallic (or SC) states do not connect to one another, then the macroscopic property must be in an insulating state.
This is consistent with the electrical resistivity data of such samples under magnetic fields as shown in Fig.~\ref{fig2}c.~\cite{Taniguchi2003}
To investigate this in detail, the magnetic field dependencies of $R(\omega)$ and $\langle\omega\rangle$ at the characteristic four points marked in Fig.~\ref{fig2} are shown in Fig.~\ref{fig4}.
The points $\alpha$ and $\beta$ have no magnetic field-dependent insulating and metallic reflectivity spectra, respectively.
This behavior is the same as that in $d[4,4]$ and $d[0,0]$, respectively (data not shown).
In comparison with these points, the magnetic field dependencies at points $\gamma$ and $\delta$ show characteristic behaviors.
At both of these points, the metallic $R(\omega)$ at 0~T changes to an insulating state at 5~T, \textit{i.e.}, the magnetic field-induced Mott transition occurs between 0 and 5~T.
In contrast to the insulating $R(\omega)$ at 10~T at point $\gamma$, $R(\omega)$ at point $\delta$ changes to the metallic state again at 10~T, {\it i.e.}, SIMT occurs at the point $\delta$.
The area in which the same field dependence as point $\delta$ appears is 6~\% of the whole measured region.

\begin{figure}[t]
\begin{center}
\includegraphics[width=7.5cm]{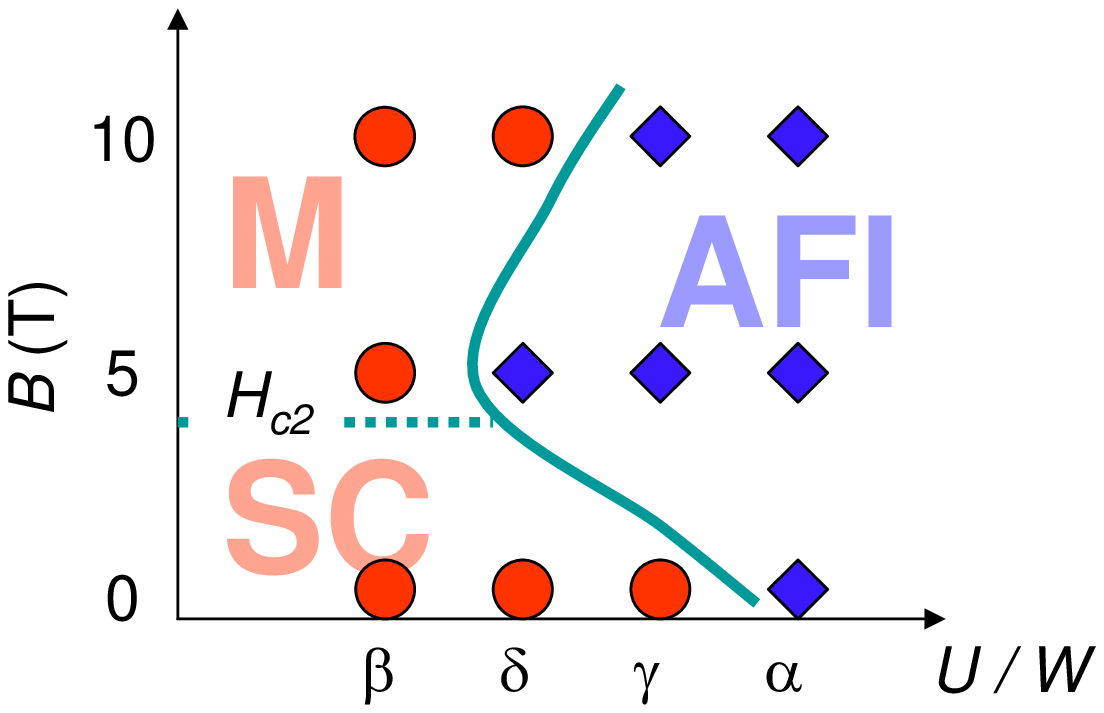}
\end{center}
\caption{
(Color online)
A conceptual $B-U/W$ phase diagram of $d[2,2]$.
The points $\alpha, \beta, \gamma$, and $\delta$ refer to the same points in Fig.~\ref{fig2}.
The solid circles and diamonds indicate the metallic (M) and insulating (AFI) points, respectively, and the phase boundary is represented by a solid line.
Below the upper critical field ($H_{c2}\sim$4~T, dashed line), the superconducting state (SC) is regarded to appear in the metallic state (M).
}
\label{fig5}
\end{figure}
The conceptual $B-U/W$ phase diagram from the magnetic field-dependent $R(\omega)$ expected from Fig.~\ref{fig4}b is shown in Fig.~\ref{fig5}.
$U/W$ is roughly proportional to $\langle\omega\rangle$ based on Ref.14.
The $\langle\omega\rangle$ at $\beta$, $\gamma$, and $\delta$ are similar to one anther at 0~T.
However, $\langle\omega\rangle$ at 10~T indicates that the $U/W$ values fall in the order of $\beta$, $\delta$, $\gamma$, and $\alpha$.
The metallic and insulating characteristics at each point under the magnetic fields are plotted with solid circles and diamonds, respectively, and the boundary is indicated by a solid line, as shown in Fig.~\ref{fig5}.
The figure indicates that the boundary between the metallic (or SC) and insulating states is not linear, but rather has a re-entrant shape.
A similar re-entrant M(SC)--I phase boundary has been observed in the $P-T$ phase diagram of $\kappa$-Cl.~\cite{Kagawa}
In one-dimensional Bechgaard salts, on the other hand, a transition from metallic to field-induced spin-density wave state has been observed.~\cite{McKernan1995}
However, to our best knowledge, our result is the first example of the magnetic field-induced SIMT.

The re-entrant phase diagram originates from the combination of the inhomogenity of the sample and the disorder of the ethylene end groups in the BEDT-TTF molecules.
The inhomogenity also affects the width of the distribution of $\langle\omega\rangle$ under the slow cooling conditions, as shown in Fig.~\ref{fig3}.
The FWHM resulting from the Gaussian fitting under slow cooling conditions is almost constant over the different magnetic fields ($46.7\pm1.2$~cm$^{-1}$ at 0~T, $58.0\pm1.9$~cm$^{-1}$ at 5~T and $51.7\pm1.4$~cm$^{-1}$ at 10~T).
Under fast cooling conditions, however, FWHM~=~$81.9\pm1.4$~cm$^{-1}$ at 0~T, $129.5\pm6.2$~cm$^{-1}$ at 5~T and $143.2\pm7.7$~cm$^{-1}$ at 10~T for a sample possessing not only the inhomogenity but also disorder effects.
The width due to the disorder effect can be evaluated to be $67.4\pm1.8$~cm$^{-1}$ at 0~T, $115.8\pm6.5$~cm$^{-1}$ at 5~T and $133.5\pm7.8$~cm$^{-1}$ at 10~T based on these values under the slow and fast cooling conditions.
The width due to the disorder effect monotonically increases with increasing magnetic field strength.
Based on these results, the disorder of the ethylene end groups must affect the electrical characteristics at high magnetic fields.
The SIMT appearing in the fast cooling experiment is regarded to have primarily originates from the disorder effect on the appropriate condition of the degree of deuteration and the inhomogenity contained in the sample itself.

\section{Conclusion}
In conclusion, the infrared reflectivity spectra and the spatial image of the center of the spectral weight of 50~\%-deuterated $\kappa$-(BEDT-TTF)$_{2}$Cu[N(CN)$_{2}$]Br were measured as functions of magnetic field and cooling rate at 5~K.
Under fast cooling conditions, the material exhibited a phase coexistence of the metallic (or superconducting) and insulating states on the sample surface in spite of the almost metallic phase in the slow cooling condition.
Parts of the superconducting state change to the insulating state at 5~T and to the metallic state at 10~T.
The superconductor--insulator--metal transition is concluded to primarily originate from the disorder effect of the ethylene end groups of the BEDT-TTF molecules combined with the degree of deuteration and the inhomogenity contained in the sample itself.

\section*{ACKNOWLEGDMENTS}
The authors would like to thank staff members at BL43IR of SPring-8 for their technical supports, and thank F.~Kagawa for fruitful discussion.
This work was performed at SPring-8 with the approval of the Japan Synchrotron Radiation Research Institute (Proposal Nos. 2003B0061, 2005A0012) and was partially supported by Grants-in-Aid for Scientific Research (Grant Nos. 16038226, 18340110) from MEXT of Japan and by the Research Foundation for Opto-Science and Technology.


\end{document}